\documentclass[aps,pra,showpacs,groupedaddress,superscriptaddress,twocolumn]{revtex4}
\usepackage{amsmath}
\usepackage{amssymb}
\usepackage{graphicx}

\begin{document}

\title{Rabi model beyond the rotating wave approximation: generation of
photons from vacuum through decoherence}
\author{T. Werlang}
\affiliation{Departamento de F\'{\i}sica, Universidade Federal de S\~{a}o Carlos, P.O.
Box 676, S\~{a}o Carlos\textit{, }13565-905, S\~{a}o Paulo\textit{, }Brazil}
\author{A. V. Dodonov}
\affiliation{Departamento de F\'{\i}sica, Universidade Federal de S\~{a}o Carlos, P.O.
Box 676, S\~{a}o Carlos\textit{, }13565-905, S\~{a}o Paulo\textit{, }Brazil}
\author{E. I. Duzzioni}
\affiliation{Centro de Ci\^{e}ncias Naturais e Humanas, Universidade Federal do ABC, Rua
Santa Ad\'{e}lia, 166, Santo Andr\'{e}, S\~{a}o Paulo, 09210-170,\textit{\ }%
Brazil}
\author{C.J. Villas-B\^{o}as}
\affiliation{Departamento de F\'{\i}sica, Universidade Federal de S\~{a}o Carlos, P.O.
Box 676, S\~{a}o Carlos\textit{, }13565-905, S\~{a}o Paulo\textit{, }Brazil}

\begin{abstract}
We study numerically the dynamics of the Rabi Hamiltonian, describing the
interaction of a single cavity mode and a two-level atom without the
rotating wave approximation, subjected to damping and dephasing reservoirs
included via usual Lindblad superoperators in the master equation. We show
that the combination of the antirotating term and the atomic dephasing leads
to linear asymptotic photons generation from vacuum. We reveal the origins of the
phenomenon and estimate its importance in realistic situations.
\end{abstract}

\maketitle

\section{Introduction}

A fundamental task in Physics is the description of the matter-light
interaction. The most simple model to deal with is the Rabi model \cite{Rabi}%
, which describes the interaction of a two-level atom with a single mode of
the quantized electromagnetic (EM) field. The Rabi Hamiltonian (RH) reads $%
(\hbar =1)$%
\begin{equation}
H=\omega a^{\dagger }a+\frac{\omega _{0}}{2}\sigma _{z}+g\left( \sigma
_{+}+\sigma _{-}\right) \left( a+a^{\dagger }\right) ,  \label{hrabi}
\end{equation}%
where $\omega $ and $\omega _{0}$ are the field and atomic transition
frequencies, respectively, and $g$ is the coupling constant (vacuum Rabi
frequency). $a$ ($a^{\dagger }$) is the annihilation (creation) operator of
the EM field, $\sigma _{z}=\left\vert e\right\rangle \left\langle
e\right\vert -\left\vert g\right\rangle \left\langle g\right\vert $ and $%
\sigma _{+}=\left\vert e\right\rangle \left\langle g\right\vert $ ($\sigma
_{-}=\sigma _{+}^{\dagger }$) are atomic operators, with $\left\vert
g\right\rangle $ and $\left\vert e\right\rangle $ denoting the ground and
excited atomic states, respectively. Although largely studied over the last
decades, up to now its exact analytical solution is lacking and only
numerical \cite{n1,n2,n3,n4} and approximate analytical solutions are
available \cite{Klim1,Pereverzev,Irish}, despite the conjecture by Reik and
Doucha \cite{Reik} that an exact solution of RH in terms of known functions
is possible. The most used analytical approach to RH is to make the rotate
wave approximation (RWA), where the antirotating term $g\left( a^{\dagger
}\sigma _{+}+a\sigma _{-}\right) $ is neglected %
, since in the weak coupling regime $g/\omega \ll 1$, small detuning $%
\left\vert \Delta \right\vert \ll \omega $ ($\Delta =\omega _{0}-\omega )$,
and weak field amplitude its contribution to the evolution of the system is
quite small \cite{Scully,Schleich}. In this limit the RH is known as Jaynes-Cummings Hamiltonian
(JCH) \cite{JC,Shore} and can be integrated exactly.

For having an exact solution, the JCH has been largely employed in Quantum
Optics, in particular in cavity quantum electrodynamics (QED) \cite%
{S298-1372,haroxo}, where the vast majority of the experiments satisfy the
required parameters regime \cite{Scully,Schleich}. JCH revealed interesting
phenomena related to the quantum nature of the light, encompassing the
granular nature of the electromagnetic field, revealed through collapse and
revivals of the atomic inversion \cite{granular}, Rabi oscillations \cite%
{rabio}, squeezing \cite{8}, non-classical states, such as Schr\"{o}dinger
cat state-like \cite{sc} and Fock states \cite{fs}, and the entanglement
between atom-atom or atom-field \cite{entanglement}. The manipulation of
atom-field interaction has been employed in the implementation of quantum
logic gates in trapped ions \cite{ql} and in cavity QED \cite{gc}, as well
as the atomic teleportation process \cite{telep}, which have contributed for
a fast development of the quantum information science \cite{infq}. Moreover,
over the past few years the JCH was experimentally investigated in solid
state cavity QED systems in the strong coupling regime using superconducting
artificial two-level atoms coupled to microwave waveguide resonators \cite%
{N431-159,N431-162,PRL96-127006} (the so called \textit{circuit} QED \cite%
{revvv}) and quantum dots coupled to photonic crystals microcavities \cite%
{N432-197,N432-200}. In fact, in circuit QED the JCH is the basic
theoretical tool for describing quantum logic gates and read-out protocols 
\cite{s3,PRA75-032329}.

The antirotating term neglected under RWA is usually wrongly interpreted as
being non-conservative \cite{nce,Scully,Schleich}, once it could allow for a
violation of the energy conservation. In fact, this term does not conserve
the \emph{total number of quanta} of the system, defined by the operator $%
N\equiv a^{\dagger }a+\sigma _{z}+\mathbb{I}$ (where $\mathbb{I}$ stands for idendity operator). However, it does not change the \emph{%
total energy} of the system, as can be easily seen through the Heisenberg
equation of motion for the total Hamiltonian operator $dH/dt=i\left[ H,H%
\right] +\partial H/\partial t=0$, once $H$ in Eq. (\ref{hrabi}) is time
independent. Therefore, the total energy of the system $\left\langle
H\right\rangle $ is conserved and no violation of physical laws occur.
Besides, recent works questioned the validity of the RWA \cite{v1,v2,v3} and
proposed alternative analytical approximate methods \cite{Irish,Pereverzev}.
Moreover, it was shown that the antirotating term is responsible for
several novel quantum mechanical phenomena, such as quantum irreversibility
and chaos \cite{chaotic,chaotic1}, quantum phase transitions \cite{qpt},
implementation of Landau-Zener transitions of a qubit in circuit QED
architecture \cite{lz,lz2}, generation of atom-cavity entanglement \cite%
{lz1,s6}, and simulation of the dynamical Casimir effect (DCE \cite{book}) in
semiconducting microcavities \cite{d1,d2,PRB72-115303} or circuit QED \cite%
{s6}.

In this work we study numerically the dynamics of the RH subjected to dissipative effects
acting on both the atom and cavity mode. These undesirable effects are
taken into account through the master equation approach \cite{Scully}, where
the Lindblad superoperators are built \textit{as usual} \cite{s1}. Without
any formal prove, we simply assume that the evolution of the density
operator of the system $\rho (t)$ is described by the master equation%
\begin{equation}
\frac{\partial \rho }{\partial t}=-i\left[ H,\rho \right] +\mathcal{L}\left(
\rho \right) ,  \label{mastereq}
\end{equation}%
where $H$ is the RH (\ref{hrabi}) and the Lindblad operator $\mathcal{L}%
\left( \rho \right) $ is given by 
\begin{equation}
\mathcal{L}\left( \rho \right) =\mathcal{L}_{a}\left( \rho \right) +\mathcal{%
L}_{d}\left( \rho \right) +\mathcal{L}_{f}\left( \rho \right) ,
\label{Lindblad}
\end{equation}%
with the standard definitions 
\begin{eqnarray}
\mathcal{L}_{a}\left( \rho \right) &=&\frac{\gamma }{2}(n_{t}+1)\left(
2\sigma _{-}\rho \sigma _{+}-\sigma _{+}\sigma _{-}\rho -\rho \sigma
_{+}\sigma _{-}\right)  \notag \\
&+&\frac{\gamma }{2}n_{t}\left( 2\sigma _{+}\rho \sigma _{-}-\sigma
_{-}\sigma _{+}\rho -\rho \sigma _{-}\sigma _{+}\right) ,  \label{La} \\
\mathcal{L}_{f}\left( \rho \right) &=&\frac{\kappa }{2}(n_{t}+1)\left(
2a\rho a^{\dagger }-a^{\dagger }a\rho -\rho a^{\dagger }a\right)  \notag \\
&+&\frac{\kappa }{2}n_{t}\left( 2a^{\dagger }\rho a-aa^{\dagger }\rho -\rho
aa^{\dagger }\right) ,  \label{Lf} \\
\mathcal{L}_{d}\left( \rho \right) &=&\gamma_{ph}\left( \sigma _{z}\rho
\sigma _{z}-\rho \right) .  \label{Ld}
\end{eqnarray}%
The superoperators $\mathcal{L}_{a}\left( \rho \right) $ and $\mathcal{L}%
_{f}\left( \rho \right) $ describe the thermal reservoirs effects (with mean photon
number $n_{t}$) on the atom and field, respectively, where $\gamma $
($\kappa $) is the atom (cavity) relaxation rate. Another source of
decoherence of the atom is the phase damping reservoir, represented by $%
\mathcal{L}_{d}\left( \rho \right) $, where $\gamma_{ph}$ is the dephasing
rate. By focusing our attention on the asymptotic \textit{photons creation
from vacuum} driven by the combination of the antirotating term $g\left(
a^{\dagger }\sigma _{+}+a\sigma _{-}\right) $ in (\ref{hrabi}) and the
atomic phase reservoir, we show that even in situations where the atom and
field are initially prepared in their individual ground states, i.e., $%
\left\vert \phi \right\rangle =\left\vert g,0\right\rangle $, where $%
\left\vert 0\right\rangle $ is the ground state of the EM field, there is an
asymptotic photons generation. This process depends on the intensity of the
atom-field coupling $g$, the detuning $\Delta$, and it
is considerably amplified when the atomic phase reservoir is predominant
over the other dissipative channels, such as atomic and field damping due to
the thermal reservoirs. The essence of the photons creation mechanism
presented here relies on the existence of the antirotating term in the RH
and the limitation\ imposed by the quantum vacuum, namely, $a\left\vert
0\right\rangle =0$. The role played by the atomic phase reservoir is just to
amplify the photons creation process through atomic decoherence. As far as
we know, this phenomenon has not been described in the literature.

This work is organized as follows. In Sec. II we study in details the
process of photons generation due to the atomic dephasing and explain its
origin. In Sec. III we give an alternative physical explanation of photons
generations by considering that the atom's dephasing is due to random shifts
of the atomic transition frequency, as usually occurs in solid state systems
(e.g. circuit QED). In Sec. IV we study the influence of damping process on
the photons creation through decoherence and estimate the net effect in
realistic situations. Finally, the section V contains discussion of results
and concluding remarks.

\section{Pure dephasing}

For convenience, from now on we set the cavity frequency to $\omega =1$.
First we integrated numerically the Eq. (\ref{mastereq}) considering just
the dephasing reservoir ($\gamma =\kappa =0$). In Figs. 1 and 2 we set the
parameters $\Delta =0$ and $g=0.1.$ These and other values of the parameters
were chosen in order to optimize the numerical calculations, and are not
related to the experimental data; nevertheless, we checked out that
qualitatively the behavior described below also holds for realistic
parameters (see Fig. 5b below). The state $\left\vert g,0\right\rangle $ is
the ground state of the JCH, so it is not coupled to other states
under the JCH dynamics. On the other hand, the antirotating term in the RH
does induce transitions to other states and, since the atom and field states
are limited from below by $\left\vert g\right\rangle $ and $\left\vert
0\right\rangle $, respectively, these transitions may only \textit{increase} 
$\left\langle n\right\rangle $. In Fig. 1a we plot the mean photon number $%
\left\langle n\right\rangle $ without dephasing ($\gamma _{ph}=0$, line 1)
as function of the dimensionless time $\tau \equiv \xi t$ with $\xi =0.1$
for the initial state $\left\vert g,0\right\rangle $, showing a bound
oscillating behavior as described in details in \cite{n2}. Fig. 1a also shows 
$\left\langle n\right\rangle $ (line 2) and the Mandel factor $q=(\left\langle \Delta
n\right\rangle ^{2}-\left\langle n\right\rangle )/\left\langle
n\right\rangle $ (line 3) for dephasing rate $\gamma _{ph}=0.1$: now $\left\langle
n\right\rangle $ increases with time, achieving asymptotically a linear
behavior, and the generated field state demonstrates a super-poissonian
behavior, $q>0$.

\begin{figure}[tbp]
\begin{center}
\includegraphics[width=.48\textwidth]{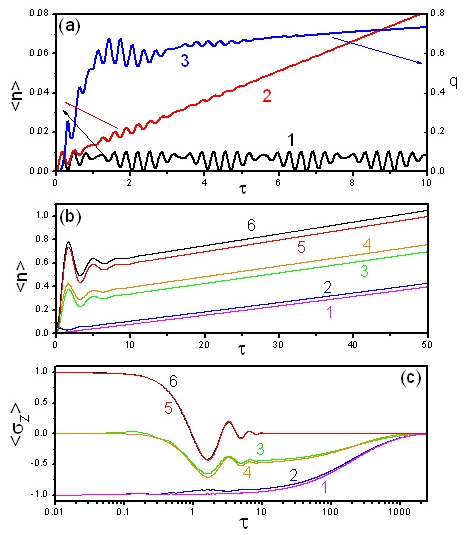} {}
\end{center}
\caption{Dynamics of the Rabi Hamiltonian for $\Delta =0$ and $g=0.1$ as
function os dimensionless time $\protect\tau =\protect\xi t$ ($\protect\xi%
=0.1$). \textbf{a}) Mean photon number $\langle n\rangle $ for initial state 
$|g,0\rangle $ without dephasing (line 1) and with dephasing $\protect\gamma %
_{ph}=0.1$ (line 2). There is photon generation from vacuum due
to atomic dephasing and the created field state is
superpoissonian, since the Mandel factor $q>0$ (line 3). \textbf{b}) $%
\langle n\rangle $ for different initial states $|\protect\phi _{i}\rangle $%
, $i=1,..,6$ (see the text), demonstrating that the asymptotic photons
generation does not depend on the initial state. \textbf{c}) Population
inversion $\langle \protect\sigma _{z}\rangle $ for the states $|\protect%
\phi _{i}\rangle $ shown in (b): as expected, $\langle \protect\sigma %
_{z}\rangle $ goes to zero asymptotically due to the decoherence.}
\end{figure}

The photons creation mechanism through atomic dephasing demonstrated in Fig.
1a is a general phenomenon and asymptotically does not depend on the initial
state of the system. In Fig. 1b we plot $\left\langle n\right\rangle $ 
\textit{versus} $\tau $ for 6 different initial states: $\left\vert \phi
_{1}\right\rangle =\left\vert g,0\right\rangle $, $\left\vert \phi
_{2}\right\rangle =\left\vert g,\alpha \right\rangle $ where $|\alpha
\rangle $ is the coherent state and we take $\left\vert \alpha \right\vert
^{2}=0.05$, $\left\vert \phi _{3}\right\rangle =[\left( \left\vert
g\right\rangle +\left\vert e\right\rangle \right) /\sqrt{2}]\otimes
\left\vert 0\right\rangle $, $\left\vert \phi _{4}\right\rangle =[\left(
\left\vert g\right\rangle +\left\vert e\right\rangle \right) /\sqrt{2}%
]\otimes \left\vert \alpha \right\rangle $, $\left\vert \phi
_{5}\right\rangle =\left\vert e,0\right\rangle $, and $\left\vert \phi
_{6}\right\rangle =\left\vert e,\alpha \right\rangle $. The basic\
difference between these initial states is the amount of quanta $%
\left\langle N_{k}\right\rangle =\langle \phi _{k}|N|\phi _{k}\rangle $: we
have $\left\langle N_{1}\right\rangle =0$, $\left\langle N_{2}\right\rangle
=0.05$, $\left\langle N_{3}\right\rangle =0.5$, $\left\langle
N_{4}\right\rangle =0.55$, $\left\langle N_{5}\right\rangle =1$ and $%
\left\langle N_{6}\right\rangle =1.05$. We see that after the transient
regime ($\tau \gtrsim 15$), which is proportional to the initial number of
quanta, all curves present the same behavior -- a linear time dependence
with the same photons creation rate. In Fig. 1c we plot the atomic
population inversion $\left\langle \sigma _{z}\right\rangle $ for these
states, showing that asymptotically $\left\langle \sigma _{z}\right\rangle $
approaches zero for any initial state. This result was expected due to the
atomic decoherence.

\begin{figure}[tbp]
\begin{center}
\includegraphics[width=.48\textwidth]{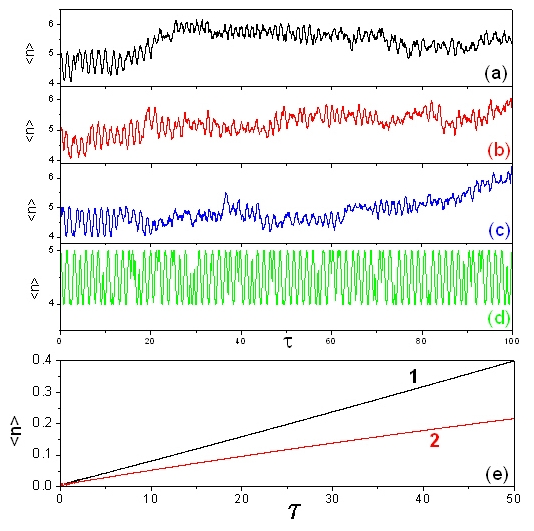} {}
\end{center}
\caption{\textbf{a}-\textbf{c}). Mean photon number for RH $\left\langle
n\right\rangle $ \textit{versus} $\protect\tau $ for three particular
trajectories, obtained using the quantum trajectories approach. The
parameters are $\Delta =0$, $g=0.1$, $\protect\gamma _{ph}=0.1$ and the
initial state is $|g,5\rangle $. $\langle n\rangle $ tends to increase with
time. \textbf{d}) $\left\langle n\right\rangle $ for a single trajectory
using JCH, showing that there is no increase of the photon number without
the antirotating terms. \textbf{e}) Asymptotic increase of mean photon
number for the RH (line 1) and the test Hamiltonian $H_{E}$ (line 2, see
text), demonstrating that the phenomenon is due to the antirotating terms
and the limitation by vacuum, $a|0\rangle =0$.}
\end{figure}

The master equation describes only the net effect of the environment on the
system. For a better understanding of the role played by the atomic
dephasing on the creation of photons we use the quantum trajectories
approach \cite{s1,Carma1} to study $\left\langle n\right\rangle $ during
individual trajectories. Here the quantum jump operator is $J\rho =\gamma
_{ph}\sigma _{z}\rho \sigma _{z}$ and the non-Hermitian Hamiltonian is $%
\widetilde{H}=H-i\left( \gamma _{ph}/2\right)\mathbb{I}$. In Fig. 2 (a)-(c) we plot $%
\left\langle n\right\rangle $ versus $\tau $ for 3 samples of individual
trajectories for the initial number state $\left\vert g,5\right\rangle $,
where we notice that in each trajectory $\left\langle n\right\rangle $ tends
to increase as the time goes on. This occurs for two reasons: (i) the
antirotating term in RH and (ii) the limitation of the cavity field from
below by the vacuum, $a\left\vert 0\right\rangle =0$.

To illustrate (i) we plot in Fig. 2d $\left\langle n\right\rangle $ obtained
via quantum trajectories approach for the JCH under the atomic dephasing,
where we see that $\left\langle n\right\rangle $ oscillates with time but
does not increase, contrary to the Figs. 2 (a)-(c). To explain the process
of photon creation, we notice that any state may be written in terms of the
basis states $\left\{ \left\vert s,n\right\rangle \right\} $ (with $%
s=\left\{ g,e\right\} $ and $\left\vert n\right\rangle $ the Fock state).
Between the jumps, the system evolves according to the Rabi Hamiltonian (\ref%
{hrabi}) (the non-hermitian part is not important due to the normalization
condition \cite{jap}) The JC term [$g(a\sigma_+ + a^\dagger\sigma_-)$]
promotes $\left\vert g,n\right\rangle \leftrightarrow \left\vert
e,n-1\right\rangle $ transitions, while the antirotating term induces $\left\vert
g,n\right\rangle \leftrightarrow \left\vert e,n+1\right\rangle $. The
combined action of both parts generates all the possible transitions.
However, the Fock states are limited from below by the vacuum state (ii), so
there are more available states $\left\vert s,m>n\right\rangle $ than $%
\left\vert s,m<n\right\rangle $ and the mean number of photons between the
jumps tends to increase. Upon a jump, the reservoir reads out the atom's
state (through the application of $\sigma _{z} $\ on the wavefunction, which
transforms $\left\vert g\right\rangle \rightarrow -\sqrt{\gamma _{ph}}%
\,\,\left\vert g\right\rangle $ and $\left\vert e\right\rangle \rightarrow 
\sqrt{\gamma _{ph}}\left\vert e\right\rangle $) so the coherence between the
states $|g\rangle $ and $|e\rangle $ is lost and subsequent evolution under $%
H$ will not bring the system to the state at the moment of the previous
jump. For this reason after each jump $\left\langle n\right\rangle $ tends
to be larger than upon the last jump, as clearly demonstrated in Figs.
2(a)-(c). After making a statistical average over many trajectories one
finds out that $\left\langle n\right\rangle $ always increases, in agreement
with Fig. 1.

One could suspect a third explanation for photons generation due to atomic
dephasing -- the different weights $\sqrt{n}$ and $\sqrt{n+1}$ arising upon
operating $a$ and $a^{\dagger }$ on the Fock state $\left\vert
n\right\rangle $. To show that this is not the case we consider the test
Hamiltonian $H_{E}$ obtained through the substitution of the operators $a$
and $a^{\dagger }$ in RH (\ref{hrabi}) by the\textit{\ exponential} \textit{%
phase operators }\cite{EPO,EPO1} $E_{-}\equiv \left( n+1\right) ^{-1/2}\,a$
and $E_{+}=E_{-}^{\dagger }$, respectively, where $E_{+}E_{-}=1-\left\vert
0\right\rangle \left\langle 0\right\vert $. By doing this we eliminate the
weight factors $\sqrt{n}$ and $\sqrt{n+1}$ from the RH, since $%
E_{+}\left\vert n\right\rangle =\left\vert n+1\right\rangle $ and $%
E_{-}\left\vert n\right\rangle =\left( 1-\delta _{n0}\right) \left\vert
n-1\right\rangle $. In Fig. 2e we plot $\,\left\langle n\right\rangle $ for
RH (line 1) and $\left\langle n_{E}\right\rangle $ obtained using $H_{E}$
(line 2) \textit{versus} $\tau $ for initial state $\left\vert \phi
\right\rangle =\left\vert g,0\right\rangle $. In both the cases there is
photon creation from vacuum, although $\left\langle n_{E}\right\rangle $
increases slower than $\left\langle n\right\rangle $. Therefore this photon
creation phenomenon is not due to the different weights $\sqrt{n}$ and $%
\sqrt{n+1}$ attributed to the operators $a$ and $a^{\dagger }$, but to the
presence of the EM vacuum state.

\begin{figure}[tbp]
\begin{center}
\includegraphics[width=.48\textwidth]{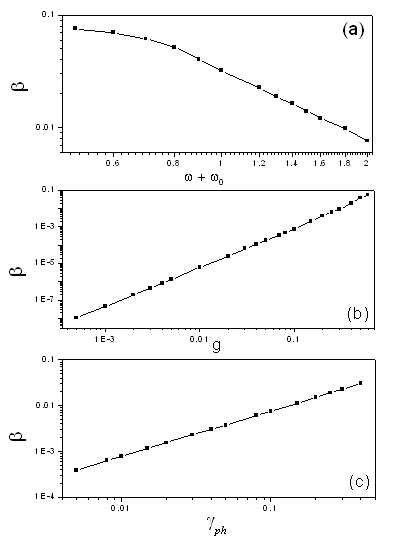} {}
\end{center}
\caption{Asymptotic photons generation rate $\protect\beta \equiv
d\left\langle n(\protect\tau )\right\rangle /d\protect\tau |_{\protect\tau %
\rightarrow \infty }$ as function of \textbf{a}) $\left( \protect\omega +%
\protect\omega _{0}\right) $ for fixed $g$ e $\protect\gamma_{ph} $; \textbf{%
b}) $g$ for fixed $\protect\gamma_{ph} $ and $\protect\omega _{0}$; \textbf{c%
}) $\protect\gamma_{ph} $ for fixed $g$ and $\protect\omega _{0}$. We
observe that $\protect\beta \sim \protect\gamma_{ph} $ and is inversely
proportional to $\protect\omega _{0}+\protect\omega $. In the weak coupling
regime $\protect\beta \sim g^{2}$.}
\end{figure}

From Fig. 1 we notice that asymptotically $\left\langle n(\tau
)\right\rangle $ increases linearly with time, so one may gain a deeper
insight into the problem by analyzing how the asymptotic photons generation
rate $\beta \equiv d\left\langle n(\tau )\right\rangle /d\tau |_{\tau
\rightarrow \infty }$ scales with $\omega +\omega _{0}$, $g$, and $\gamma
_{ph}$ (here we fix $\omega $ and vary other parameters). In Fig. 3a we plot 
$\beta $ \textit{versus} $\left( \omega +\omega _{0}\right) $ for fixed $g$
e $\gamma _{ph}$, where we see that $\beta $ increases when $\omega _{0}$
decreases, so $\beta $ is inversely proportional to $\left( \omega +\omega
_{0}\right) $. Fig 3b shows the dependence of $\beta $ on $g$ for fixed $%
\gamma _{ph}$ and $\omega _{0}$: for $g\ll 1$, the \textit{weak coupling
regime}, the analysis of the curve shows that $\beta \sim g^{2}$, however
such a dependence is modified for large values of $g$. Finally, Fig.3c shows
that $\beta \sim \gamma _{ph}$ for fixed $g$ and $\omega _{0}$, in agreement
with the quantum trajectories approach: indeed, larger $\gamma _{ph}$
implies larger jump probability and, consequently, in average $\left\langle
n\right\rangle $ increases faster with time. A quantitative analysis of $%
\beta $ will be presented elsewhere.

\section{Random frequency fluctuations}

One of the origins of dephasing from the physical point of view are the
random shifts of the atomic transition frequency $\omega _{0}$ due to the
interaction with the environment \cite{s1,s2}. Indeed, 1/$f$ noise in the
bias controlling the atomic transition frequency is the dominant source of
decoherence in superconducting artificial atoms (qubits) \cite{s3,s4,s5,phh}%
. To investigate the effect of such a noise on the dynamics of the
atom-cavity system, we integrated numerically the master equation for the RH
(neglecting the damping and dephasing) assuming that $\omega _{0}$ has
stochastic fluctuations. Our goal is to show that, when averaged over
ensemble, this source of decoherence does asymptotically generate photons
from vacuum, since its mathematical description is given by the dephasing
Lindblad superoperator in the master equation we studied above.

As a simple model we considered time-dependent $\omega _{0}(t)$ 
\begin{equation}
\omega _{0}\left( t+dt\right) =\omega _{0}\left( t\right) +\left\{ 
\begin{array}{c}
0.1\varepsilon xr\text{ if }\omega _{0}(t)<\Omega _{0}-0.8\,\varepsilon  \\ 
-0.1\varepsilon x\,r\text{ if }\omega _{0}(t)>\Omega _{0}+0.8\,\varepsilon 
\\ 
0.1\varepsilon x\,\left( r-1/2\right) \text{ otherwise}%
\end{array}%
\right. ,  \label{of}
\end{equation}%
where $\Omega _{0}\equiv \omega _{0}\left( t=0\right) $ is the mean atomic
transition frequency, $r\in (-1,1)$ is a random number, $\varepsilon \ll 1$
is the maximum shift of the atom frequency and $dt$ is the simulation unit
step, $g\,dt\ll 1$. Here $x$ is related to the `frequency' of the noise:
qualitatively, for $x\ll 1$ we have `low frequency' noise, and in the
opposite limit we have `high frequency' noise.

We assumed the initial state $|g,0\rangle $ and calculated the ensemble
average of the mean photon number $\left\langle n\right\rangle _{av}$ and
the probability of exciting the atom $P_{e}$ using the parameters $\Omega
_{0}=1$, $g=6\cdot 10^{-2}$, $\varepsilon =g$. We considered three examples
of noise whose spectra are shown in Fig. 4a: $x=1$ corresponds to the `low
frequency' noise (line 1) and $x=6$ is our reference to the `high frequency'
noise (line 3), while $x=3$ is a case in between we call `middle frequency'
noise (line 2, data not shown in Fig. 1a since it lies between lines 1 and
3). First, the Fig. 4b shows three samples of $\left\langle n\right\rangle $
obtained for single runs of simulation for the `high frequency' noise -- one
may see random rises and falls of photon number, however in average $%
\left\langle n\right\rangle $ increases. The Figs. 4c and 4d show $%
\left\langle n\right\rangle _{av}$ and $P_{e}$, respectively, obtained after
averaging out many runs of simulations for the three kinds of noise. We see
that in average there is a growth of both $\left\langle n\right\rangle _{av}$
and $P_{e}$; moreover, the growth of $\left\langle n\right\rangle _{av}$ is
approximately linear in time, in agreement with our previous results.

\begin{figure}[tbp]
\begin{center}
\includegraphics[width=.48\textwidth]{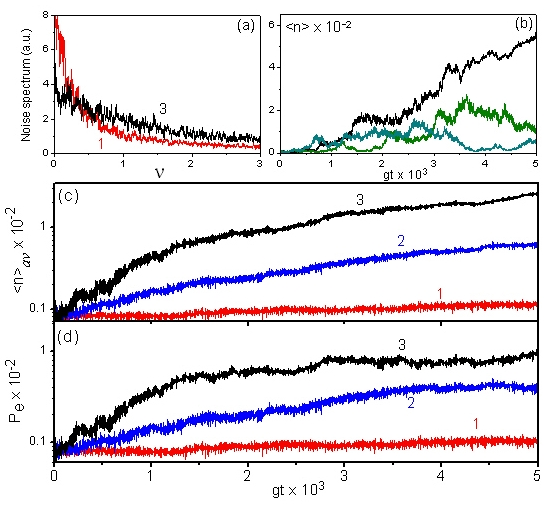} {}
\end{center}
\caption{Simulation of atomic dephasing via random atomic frequency
fluctuations for parameters $\Omega _{0}=1,$ $g=6\cdot 10^{-2}$, $\protect%
\varepsilon =g$. \textbf{a}) Spectrum of the frequency noise $\protect\omega %
_{0}(t)-\Omega _{0}$. Line 1 (red) corresponds to the `low-' and line 3
(black) -- to the 'high-' frequency noises. \textbf{b}) $\left\langle
n\right\rangle $ for three individual runs of simulation using high
frequency noise. \textbf{c}) $\left\langle n\right\rangle _{av}$ averaged
out over many runs of simulation, showing photons growth, dependent on the
noise frequency. Here line 2 (blue) is the `middle-frequency' noise. \textbf{%
d}) $P_{e}$ averaged out over many simulations. These curves agree
qualitatively with the results obtained above using the master equation
approach.}
\end{figure}

From Fig. 4c we see that the photon generation rate is higher for higher
frequency noise. One may understand qualitatively such a behavior as
follows. The dynamics of the RH with  externally prescribed non-random $%
\omega _{0}\left( t\right) $ allows for the coherent generation of both EM
and atomic \emph{real} excitations from vacuum for, e.g., periodic \cite%
{d1,d2,PRB72-115303,s6} or linear \cite{lz,lz1} time-dependence of $\omega
_{0}\left( t\right) $. Moreover,\ the photon creation rate strongly depends
on the shape of $\omega _{0}\left( t\right) $ \cite{lz} and, in the periodic
case, on the periodicity of the modulation of $\omega _{0}\left( t\right) $ 
\cite{d1,s6}. The Fourier transform (Fig. 4a) of the noise contains the
`resonant' frequencies ($\nu \sim 2$ \cite{d1,s6}), with respective weights,
for which photon generation occurs in the periodic case, so in average one
expects a slow \textit{incoherent} photon creation from vacuum due to these
components in the noise spectrum. For the high frequency noise there are
more resonant frequencies in the noise spectrum and/or their weight is
larger compared to the low-frequency noise, so the photon growth rate is
higher, in agreement with Fig. 4c.

Therefore, one of the physical origins of the photon creation through
decoherence in cavity QED are the random fluctuations of the atomic
transition frequency giving rise to effective time-dependent RH, for which
photons are created from vacuum for the `resonant' frequencies \cite{d1,s6}
present in the noise spectrum.

\section{Dephasing plus relaxation}

\begin{figure}[tbp]
\begin{center}
{\includegraphics[width=.48\textwidth]{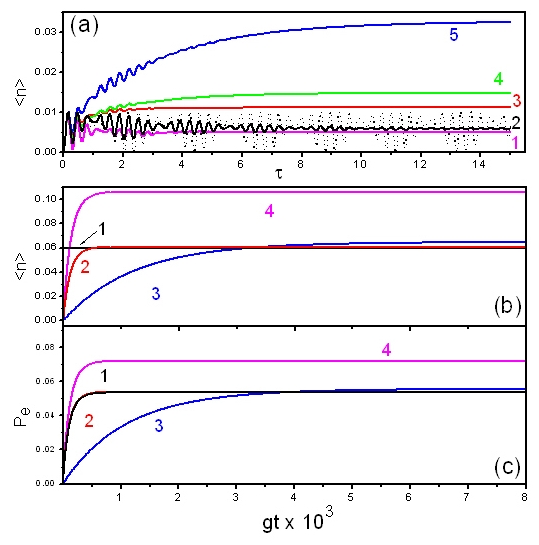}} {}
\end{center}
\caption{\textbf{a}) $\left\langle n\right\rangle $ \textit{versus} $\protect%
\tau $ using the master equation (\protect\ref{mastereq}) for initial state $%
|g,0\rangle $ with parameters $\Delta =0$, $g=0.1$, $n_{t}=0$ and decay
rates $\left( \protect\gamma_{ph} ,\protect\gamma ,\protect\kappa \right) $.
Dotted line: $(0,0,0)$, line 1: $\left( 0,0,1\right) \cdot 10^{-1}$, line 2: 
$\left( 0,1,0\right) \cdot 10^{-1}$, line 3: $\left( 1,1,1\right) \cdot
10^{-1}$, line 4: $\left( 1,0,1\right) \cdot 10^{-1}$, line 5: $\left(
1,1,0\right) \cdot 10^{-1}$. \textbf{b}) $\left\langle n\right\rangle $ 
\textit{versus }$gt$ for initial state $|g,0\rangle $ using circuit QED
parameters $\Delta =0$, $g=2\cdot 10^{-2}$, $n_{t}=6\cdot 10^{-2}$ and
distinct decay rates (line 1 denotes the thermal photon number $n_{t}$).
Line 2: current parameters $\left( 2,3,0.4\right) \cdot 10^{-4}.$ Line 3:
expected future scenario $\left( 2,3,0.4\right) \cdot 10^{-5}$. Line 4:
highly biased noise $\left( 200,3,0.4\right) \cdot 10^{-4}$. For current
parameters there is no observable difference between $n_{t}$ and $%
\left\langle n\right\rangle $, however in the future or in very noisy
environments $\left\langle n\right\rangle $ may become significantly larger
than $n_{t}$ due to photon creation through decoherence. \textbf{c)} $P_{e}$
corresponding to the parameters in (b); $P_{e}$ resembles the behavior of $%
\left\langle n\right\rangle .$}
\end{figure}

In realistic situations, besides the dephasing reservoir there are other
important error sources (environments) acting on the system, e.g.,
thermal reservoirs. When other reservoirs are present, there is a
competition between photon creation due to the atomic dephasing and photon
losses due to the damping. In order to see this effect, we plot in Fig. 5a $%
\left\langle n\right\rangle $ \textit{versus} $\tau $ for different decay
rates in Eq. (\ref{mastereq}), for the parameters $\Delta =0$, $g=0.1$, $%
n_{t}=0$ and the initial state $\left\vert g,0\right\rangle $. The effect of
temperature ($n_{t}>0$) is just to shift the curves upward. The dotted line
shows $\left\langle n\right\rangle $ in absence of any reservoir. When the
atomic phase reservoir is switched off, the curve 1 with dissipation
parameters $\left( 0,0,1\right) \cdot 10^{-1}$ [we use notation $\left(
\gamma _{ph},\gamma ,\kappa \right) $] and 2 with $\left( 0,1,0\right) \cdot
10^{-1}$ show that for large times $\left\langle n\right\rangle $ is smaller
than in the cases where the phase reservoir is switched on, as shown in the
curves 3 with $\left( 1,1,1\right) \cdot 10^{-1}$, 4 with $\left(
1,0,1\right) \cdot 10^{-1}$ and 5 with $\left( 1,1,0\right) \cdot 10^{-1}$.
Even in situations in which the environment starts at $T=0K$, in the
asymptotic limit the system behaves as being subjected to an effective
reservoir with $\tilde{n}_{t}>0$, since $\lim_{\tau \rightarrow \infty }$ $%
\left\langle n(\tau )\right\rangle >0$. The number of effective reservoir
photons $\tilde{n}_{t}$ increases when the atomic phase reservoir is present
(see curves 3, 4, and 5) and decreases when the atomic and field thermal
reservoirs are predominant. However, the effective reservoir cannot be
compared to the usual thermal reservoir, since the statistics of created
field state is quite different from the thermal state statistics. A similar
conclusion was drawn in \cite{v1}, where the authors considered the master
equation (\ref{mastereq}) at zero temperature with $\gamma _{ph}=\gamma =0$, 
$\kappa \neq 0$ and showed that the antirotating part of the RH gives rise
to a thermal-like term in the effective master equation, although it cannot
be interpreted as an interaction with a thermal bath at a certain
temperature.

In fig. 5b we consider current experimental parameters taken from recent
circuit QED experiments \cite{N445-515} $\Delta =0$, $g=2\cdot 10^{-2}$, $%
n_{t}=6\cdot 10^{-2}$ . The line 1 shows the thermal photon number $n_{t}$
and the line 2 shows $\left\langle n\right\rangle $ obtained via master
equation (\ref{mastereq}) for current dissipation rates: $\kappa =4\cdot
10^{-5}$, $\gamma =3\cdot 10^{-4}$, $\gamma _{ph}=2\cdot 10^{-4}$. There is
no visible deviation of $\left\langle n\right\rangle $ from $n_{t}$ for the
current temperatures, even if the detuning is increased to $\Delta =-0.2$
(data not shown). Next we consider the scenario expected in the future,
where the damping losses are suppressed one order of magnitude, $\kappa
=4\cdot 10^{-6}$, $\gamma =3\cdot 10^{-5}$, but the dephasing rate remains
the same, $\gamma _{ph}=2\cdot 10^{-4}$. In this case (line 3), $%
\left\langle n\right\rangle $ slightly deviates from the thermal photon
number $n_{t}$ and such an effect could be observed in very accurate
measurements. Finally, we take the current values of damping rates, $\kappa
=4\cdot 10^{-5}$, $\gamma =3\cdot 10^{-4}$, and consider a highly biased
noise \cite{phh} with the dephasing rate two order of magnitudes larger than
the best one available today, $\gamma _{ph}=2\cdot 10^{-2}$ (line 4). In
this case, $\left\langle n\right\rangle $ equals almost twice the thermal
photons number due to the phenomenon of photon creation through
decoherence. %

Finally, in Fig. 5c we plot $P_{e}$ corresponding to the parameters of Fig.
5b, where the line 1 shows $P_{e}^{RWA}$ obtained for current dissipation
parameters using RWA, which nearly does not depend on the relaxation rates.
The behavior of $P_{e}$ resembles the one of $\left\langle n\right\rangle $
and indicates that $P_{e}$ increases due to the combined action of dephasing
and the antirotating term, although for current parameters this phenomenon
is insignificant. However, for a large $\gamma _{ph}$ (line 4) $P_{e}$ is
substantially higher than $P_{e}^{RWA}$, demonstrating that large dephasing,
besides decoherence, may also induce bit flip error.

\section{Discussion and concluding remarks}

We studied numerically the dynamics of the Rabi Hamiltonian subjected to
dissipative losses, assuming \emph{ad hoc} that one may describe the
dissipative dynamics of the RH by using the standard master equation with
usual damping and dephasing Lindblad superoperators. We found out that the
atomic dephasing, when combined with the antirotating term in the RH,
induces photon creation from vacuum. The physical interpretation of the
phenomenon was given using two alternative approaches: 1) the quantum
trajectories approach based on quantum jumps and 2) microscopic \emph{ad hoc}
model of dephasing based on stochastic oscillations of the atomic transition
frequency (as occurs in solid state cavity QED). We showed that the photon
creation through atomic decoherence is suppressed in the presence of damping
mechanisms, and estimated the magnitude of this phenomenon using currents
experimental values of parameters, noting that the phenomenon might become
relevant in future experiments.

We \emph{do not} have a formal proof that the master equation (\ref{mastereq}%
) we used throughout this work is valid for the Rabi Hamiltonian in the
strong atom-cavity coupling regime, since we assumed the Lindbladian
dissipative superoperators without a microscopic deduction of the master
equation. Nevertheless, the results obtained here are relevant for two
reasons. First, if the master equation (\ref{mastereq}) is indeed
applicable, it says that decoherence induces photons generation from vacuum,
and such an effect may become relevant in future experiments with lower
temperatures and lower damping rates. Second, if the master equation (\ref%
{mastereq}) turns out to be non-applicable to this problem, it will call
attention that there is an incompatibility between the standard Lindbladian
superoperators and antirotating terms in the Rabi Hamiltonians. In any
case, more investigation of the problem is needed, since up to now the
theoretical and numerical investigations were mainly concerned about the
role of the RH antirotating term in the closed system dynamics, while our
study points out novel important effects of the antirotating term in open
systems.

\begin{acknowledgments}
The authors would like to thank the Brazilian agencies CNPq (T.W. and
C.J.V-B), FAPESP \#04/13705-3 (AVD), and UFABC (EID). This work was
supported by Brazilian Millennium Institute for Quantum Information and
FAPESP \#2005/04105-5.
\end{acknowledgments}


\begin{thebibliography}{99}
\bibitem{Rabi} I. I. Rabi, Phys. Rev. \textbf{49}, 324 (1926); \textbf{51},
652 (1937).

\bibitem{n1} C. Emary, Int. J. Mod. Phys. B. \textbf{17}, 5477 (2003).

\bibitem{n2} R. F. Bishop and C. Emary, J. Phys. A: Math. and General 
\textbf{34}, 5635 (2001).

\bibitem{n3} R. F. Bishop, \emph{et al.}, Phys. Lett. A \textbf{254}, 215
(1999); Phys. Rev. A \textbf{54}, R4657 (1996).

\bibitem{n4} V. Fessatidis, J. D. Mancini, and S. P. Bowen, Phys. Lett. A 
\textbf{297}, 100 (2002).

\bibitem{Pereverzev} A. Pereverzev and E. R. Bittner, Physical Chemistry
Chemical Physics \textbf{8}, 1378 (2006).

\bibitem{Klim1} N. Debergh and AB Klimov, Int. J. Mod. Phys. A \textbf{16},
4057 (2001).

\bibitem{Irish} E. K. Irish, Phys. Rev. Lett. \textbf{99}, 173601 (2007).

\bibitem{Reik} H. G. Reik and M. Doucha, Phys. Rev. Lett. \textbf{57}, 787
(1986); H. Reik, \textit{et al}. J. Phys. A \textbf{20}, 6327 (1987).

\bibitem{Scully} M. O. Scully and M. S. Zubairy, Quantum Optics, Cambridge
University Press, 1997.

\bibitem{Schleich} W. P. Schleich, Quantum Optics in Phase Space, WILEY-VCH
Verlag, Berlin, 2001.

\bibitem{JC} E. T. Jaynes and F. W. Cummings, Proc. IEEE \textbf{51}, 89
(1963).

\bibitem{Shore} B. W. Shore and P. L. Knight, J. Mod. Opt. \textbf{40}, 1195
(1993).

\bibitem{S298-1372} H. Mabuchi and A. C. Doherty, Science \textbf{298}, 1372
(2002).

\bibitem{haroxo} J. M. Raimond, M. Brune, and S. Haroche, Rev. Mod. Phys. 
\textbf{73}, 565 (2001).

\bibitem{granular} G. Rempe and H.Walther, Phys. Rev. Lett. \textbf{58}, 353
(1987).

\bibitem{rabio} M. Brune, \emph{et al.}, Phys. Rev. Lett. \textbf{76}, 1800
(1996).

\bibitem{8} J. R. Kuklinski and J. L. Madajczyk, Phys. Rev. A \textbf{37},
3175 (1988);

\bibitem{sc} M. Brune, \emph{et al.}, Phys. Rev. A \textbf{45}, 5193 (1992).

\bibitem{fs} M. Weidinger, B. T. H. Varcoe, R. Heerlein, and H. Walther,
Phys. Rev. Lett. \textbf{82}, 3795 (1999).

\bibitem{entanglement} S. J. D.Phoenix, and P. L. Knight, Phys. Rev. A 
\textbf{44}, 6023 (1991); Muhammed Y\"{o}na\c{c}, Ting Yu and J. H. Eberly,
J. Phys. B: At. Mol. Opt. Phys. \textbf{39,} S621-S625 (1996).

\bibitem{ql} J. I. Cirac and P. Zoller, Phys. Rev. Lett. \textbf{74}, 4091
(1995).

\bibitem{gc} T. Pellizzari, S. A. Gardiner, J. I. Cirac, and P. Zoller,
Phys. Rev. Lett. \textbf{75}, 3788 (1995); T. Sleator and H. Weinfurter,
Phys. Rev. Lett. \textbf{74}, 4087 (1995).

\bibitem{telep} S-B. Zheng and G-C. Guo, Phys. Rev. Lett. \textbf{85}, 2392
(2000); M.D. Barrett, et. al., Nature \textbf{429}, 737 (2004).

\bibitem{infq} M. A. Nielsen and I. L. Chuang, Quantum Computation and
Quantum Information, Cambridge University Press, 2000.

\bibitem{N431-159} I. Chiorescu \emph{et al.}, Nature \textbf{431}, 159
(2004).

\bibitem{N431-162} A. Wallraff \emph{et al.}, Nature \textbf{431}, 162
(2004).

\bibitem{PRL96-127006} J. Johansson \emph{et al.}, Phys. Rev. Lett. \textbf{%
96}, 127006 (2006).

\bibitem{revvv} A. Zagoskin and A. Blais, Physics in Canada \textbf{63}, 215
(2007).

\bibitem{N432-197} J. P. Reithmaier \emph{et al.}, Nature \textbf{432}, 197
(2004).

\bibitem{N432-200} T. Yoshie \emph{et al.}, Nature \textbf{432}, 200 (2004).

\bibitem{s3} A. Blais et al., Phys. Rev. A \textbf{69}, 062320 (2004).

\bibitem{PRA75-032329} A. Blais \emph{et al.}, Phys. Rev. A \textbf{75},
032329 (2007).

\bibitem{nce} C. Gerry and P. Knight, Introductory Quantum Optics, Cambridge
University Press, 2005.

\bibitem{v1} A. B. Klimov, J. L. Romero, and C. Saavedra, Phys. Rev. A 
\textbf{64}, 063802 (2001).

\bibitem{v2} J. Larson, Phys. Scr. \textbf{76}, 146 (2007).

\bibitem{v3} G. Berlin and J. Aliaga, J. Opt. B: Quant. Semiclass. Opt. 
\textbf{6}, 231 (2004).

\bibitem{chaotic} L. Bonci, R. Roncaglia, B. J. West, and P. Grigolini,
Phys. Rev. Lett. \textbf{67}, 2593 (1991).

\bibitem{chaotic1} C. Emary and T. Brandes, Phys. Rev. Lett. \textbf{90},
044101 (2003).

\bibitem{qpt} C. Emary and T. Brandes, Phys. Rev. E \textbf{67}, 066203
(2003).

\bibitem{lz} K. Saito \emph{et al.}, Europhys. Lett. \textbf{76}, 22 (2006).

\bibitem{lz2} K. Saito, M. Wubs, S. Kohler, Y. Kayanuma, and P. H\"{a}nggi,
Phys. Rev. B \textbf{75}, 214308 (2007).

\bibitem{lz1} M. Wubs M, S. Kohler S, and P. Hanggi, Physica E -
Low-Dimensional Systems and Nanostructures \textbf{40}, 187 (2007). 

\bibitem{s6} A. V. Dodonov, \emph{et al.}, unpublished.

\bibitem{book} V. V. Dodonov, Modern Nonlinear Optics, second ed, volume 
\textbf{119}, 309 (2001). 

\bibitem{d1} C. Ciuti and I. Carusotto, J. Appl. Phys. \textbf{101}, 081709
(2007). 

\bibitem{d2} S. De Liberato, C. Ciuti, and I. Carusotto, Phys. Rev. Lett. 
\textbf{98}, 103602 (2007). 

\bibitem{PRB72-115303} C. Ciuti, G. Bastard, and I. Carusotto, Phys. Rev. B 
\textbf{72}, 115303 (2005). 

\bibitem{s1} H. Carmichael, An open system approach to quantum optics,
Springer-Verlag, 1993.

\bibitem{Carma1} M. B. Plenio and P. L. Knight, Rev. Mod. Phys. \textbf{70},
101 (1998) .

\bibitem{jap} H. Goto and K. Ichimura, Phys. Rev. A \textbf{72}, 054301
(2005).

\bibitem{EPO} F. London, Z. Phys. \textbf{37}, 915 (1926); \textbf{40}, 193
(1927).

\bibitem{EPO1} L. Susskind and J. Glogower, Physics \textbf{1}, 49 (1964).

\bibitem{s2} Yu. Makhlin, G. Sch\"{o}n, and A. Shnirman, Rev. Mod. Phys. 
\textbf{73}, 357 (2001).

\bibitem{s4} G. Ithier \textit{et al}., Phys. Rev. B \textbf{72}, 134519
(2005).

\bibitem{s5} P. J. Leek \textit{et al}., Science \textbf{318}, 1889 (2007).

\bibitem{phh} P. Aliferis, \emph{et al.}, preprint at arXiv: 0806.0383.

\bibitem{N445-515} D. I. Schuster \emph{et al.}, Nature \textbf{445}, 515
(2007).
\end{thebibliography}
\end{document}